**Light emission from gold nanoparticles under ultrafast near-infrared excitation: thermal emission, inelastic light scattering or multiphoton luminescence?**

Lukas Roloff, Philippe Klemm, Imke Gronwald, Rupert Huber, John M. Lupton, Sebastian Bange[*)]

Institut für Experimentelle und Angewandte Physik, Universität Regensburg, 93051 Regensburg, Germany

**Abstract:** Gold nanoparticles emit broad-band upconverted luminescence upon irradiation with pulsed infrared laser radiation. Although the phenomenon is widely observed, considerable disagreement still exists concerning the underlying physics – most notably over the applicability of concepts such as multiphoton absorption, inelastic scattering, and interband and intraband electronic transitions. Here, we study single particles and small clusters of particles by employing a spectrally resolved power-law analysis of the irradiation-dependent emission as a sensitive probe of these physical models. Two regimes of emission are identified: at low irradiance levels of kW/cm², the emission follows a well-defined integer-exponent power law suggestive of a multiphoton process. However, at higher irradiance levels of several kW/cm², the nonlinearity exponent itself depends on the photon energy detected, a tell-tale signature of a radiating heated electron gas. We show that in this regime, the experiments are incompatible with both interband transitions and inelastic light scattering as the cause the luminescence, while they are compatible with the notion of luminescence linked to intraband transitions.

The absorption of laser radiation in metal nanostructures can be understood as a collective plasmonic response of the electron gas, followed by dephasing and electron-electron scattering, and results in a high-temperature Fermi-Dirac distribution in the conduction band.[1] The subsequently coupled evolution of electronic and lattice temperatures on the timescale of a few picoseconds, known as the two-temperature model, is equally applied in the fields of materials science[2], plasmonics[3], and ultrafast surface chemistry[4]. Although the inverse process to absorption – light emission from a hot conduction-band electron system through intraband transitions – was described in the 1970s,[5] it remains less well understood. With the plasmonic field enhancement provided by nanostructured metal surfaces, broad-band luminescence phenomena gained scientific attention mainly out of concern for their appearance as a background signal in surface-enhanced Raman scattering (SERS). Such light emission spans more than one eV in photon energy[6-8] and was shown to emanate from the metal particles themselves instead of the "hot spot" gaps that are commonly associated with the plasmonic field enhancement and the Raman signal[9-10]. The phenomenon is much stronger under pulsed excitation and thus highly relevant for future applications of ultrafast SERS techniques.[11] Recently, convincing evidence has been provided for the case of near-infrared c.w. excitation that suggests that the luminescence should be attributed to photons scattering off conduction-band electrons.[12-14] Such inelastic light scattering – or electronic Raman scattering – has been a valuable tool in the study of correlated electron systems such as superconductors.[15] It is disputed, though, if the background signal in c.w. experiments can be fully accounted for by this mechanism,[16] and the model so far failed to fully account for the Stokes side of the spectrum observed in pulsed laser experiments[17]. While direct interband transitions are possible under visible-light excitation,[18] they are out of reach of single-photon absorption for excitation with near-infrared lasers at around 1.5 eV. Considering that intra-conduction-band transitions are both symmetry forbidden and non-conserving in electron momenta, no obvious candidate for a first-order luminescent interaction thus exists in this case.[19] So far, the main agreement across different studies has been about the nonlinearity of the light emission intensity with respect to the pulsed infrared excitation power.[7,8,20-28] This effect has been taken as a sign of genuine multiphoton absorption processes, i.e. one electron absorbing multiple photons.[25,28] In the light of strong electron-electron scattering in the conduction band, absorption has also been discussed in terms of sequential intraband transitions, possibly leading to luminescence due to recombination of electrons from the heated *sp*-conduction band with holes in the lower-lying *d*-bands.[26,27] Absorption-induced heating of the conduction band introduces an effective nonlinear response into the system which can only be distinguished from multiphoton absorption by careful analysis of the excitation pulse-length dependence[17] and through excitation by multiple closely spaced laser pulses[29]. Since the emission spectrum is strongly influenced by the antenna effect provided by the longitudinal and transverse particle plasmon oscillation modes,[28] direct emission spectroscopy does not yield useful insight into the physical origin of the luminescence. We recently introduced the idea of studying the nonlinearity of metal luminescence excited by ultrafast infrared laser pulses as a function of the energy of the emitted photons, rather than in a spectrally integrated form.[30] Looking only at the relative luminescence changes as a function of irradiation, the method is insensitive to linear scaling of the emission by optical antenna effects. For nanoscopically rough silver and gold surfaces, the nonlinearity could be described by a power-law exponent that featured a nearly linear dependence on the energy of the emitted photons. This was shown to be in close agreement with a model based on conduction-band heating being driven by photon absorption and luminescence occurring from intraband transitions within the energy distribution of the hot electron gas.[30] Here, we apply this spectrally resolved power-law analysis to study luminescence from individual and clustered gold nanoparticles under pulsed infrared laser excitation.

Gold nanorods with 38 nm length and 10 nm diameter capped by cetyltrimethylammonium bromide (CTAB) and dispersed in water were purchased from Creative Diagnostics. After sufficient dilution with ultrapure water, the particles were deposited by spin-coating directly onto thoroughly cleaned microscopy glass cover slips or onto glass cover slips (Fisher Scientific) covered by 100 nm of transparent indium-tin oxide (ITO, Evaporated Coatings Inc.), as depicted schematically in Figure 1a. Prior to optical studies, nanoparticles on top of the conductive ITO layer were imaged with a scanning electron microscope (Zeiss Supra 40, using in-lens detection) to determine the position of individual particles, see Figure 1b. Spatial correlation between fluorescence and electron microscopy was achieved through a reference pattern of micrometer-sized holes in the ITO layer drilled by fs-pulsed laser ablation prior to sample preparation. For optical studies, a custom laser microscopy setup was used, which provided through-substrate laser excitation and fluorescence collection by an oil-immersion microscope objective (numerical aperture 1.49, Olympus). The excitation was provided by a mode-locked Ti:Sapphire laser (SpectraPhysics) operating at a photon energy of 1.62 eV (wavelength of 766 nm), and emitting laser pulses of approximately 72 fs length at a repetition rate of 80 MHz. The excitation wavelength was chosen to match the wavelength of peak nanoparticle absorbance, plotted in Figure 1c, which is associated with their longitudinal plasmon mode. Fluorescence was collected through either a 1.5 eV low-pass filter (Stokes) or a 1.8 eV high-pass filter (anti-Stokes) to suppress scattered laser radiation, and imaged onto a cooled sCMOS camera (Hamamatsu) for spatial luminescence maps, or onto an imaging spectrometer with a cooled CCD camera (Princeton Instruments) for the collection of emission spectra.

Figure 1b shows representative fluorescence maps collected in the anti-Stokes spectral range at photon energies above 1.8 eV, using circularly polarized wide-field laser excitation, superimposed on the SEM topography. Magnified SEM images show that both single nanoparticles (site A) as well as small clusters of nanoparticles (site C) contribute to individual, resolution-limited luminescence sites. Figure 1d shows individual luminescence spectra from these sites under strong, focused laser excitation for both the anti-Stokes as well as Stokes (< 1.5 eV) spectral range. Under these conditions, anti-Stokes spectra appear virtually background-free and cover the spectral range up to a photon energy $h\nu$ of 3.2 eV. In contrast, Stokes luminescence measured from sites A and C cannot easily be distinguished from the luminescence measured either from a close-by particle-free sample area (site B) or from the surface of a clean glass/ITO substrate. Figure 1e shows the integrated luminescence intensity $\phi(E)$ as a function of irradiance $E$ in three indicated narrow spectral ranges. The irradiance was kept below the threshold for photomodification, so that no significant change of luminescence behavior is seen in upward and downward sweeps of the irradiation level. In all cases, a power-law-type behavior $\phi(E) \propto E^p$ prevails, where the power-law exponent $p$ is found from a linear fit to the double-logarithmic representation of $\phi(E)$. For the single particle (site A), the exponent $p$ rises with photon energy from $p = 1.6$ at $h\nu = 1.3$ eV to $p = 2.7$ at 2.0 eV and $p = 3.5$ at 2.6 eV. The corresponding values for the particle aggregate (site C) are $p = 1.8$ at 1.3 eV, $p = 2.7$ at 2.0 eV and $p = 3.5$ at 2.6 eV. For the particle-free site B we find $p = 1.2$ at 1.3 eV, and for the bare substrate $p = 1.0$, also at 1.3 eV. In both cases no anti-Stokes emission is observed. Errors in $p$ were below 0.2 for all fits. Although correcting for background luminescence is challenging under focused excitation and the Stokes spectrum of particle sites is masked by emission originating directly from the substrate, the irradiance dependence of the emission is clearly superlinear, suggesting that the particle also contributes measurably to the Stokes emission. The power-law exponent is significantly higher for the anti-Stokes luminescence and further increases along with the energy of the detected photons. Non-integer values of the exponent $p$ and the dependence of $p$ on the emission photon energy cannot be easily reconciled with the common notion of nonlinear luminescence being driven by multiphoton interband transitions[25].

Spectrally integrated measurements have previously pointed to the emergence of non-integer power-law behavior, although the deviations from integer behavior were mostly ignored.[24] Experiments with double-pulse excitation have clearly shown that the nonlinearity is higher with respect to varying the power of the first laser pulse than that of the second and that the correlation time scales are on the order of the electron-phonon scattering time of a few ps.[29] Indeed, the observations are readily explained by assuming that the interaction with the laser pulses heats the conduction-band electron gas to an effective temperature $T_e$ well above the initial conditions set by the lattice temperature, and that luminescence is correlated with the occurrence of high-energy intraband transitions.[30] The spectral dependence of the power-law exponent $p(h\nu)$ is then merely a result of the irradiation-dependent blue shift of the energy distribution as is well known, e.g. for blackbody radiation. For bulk metals, the electronic temperature $T_e$ depends on the irradiance $E$ as $T_e^a \propto E$. Conventionally, an exponent $a = 2$ is expected from the electronic heat capacity,[31] although this exponent for the *effective* temperature ultimately depends on the time evolution of the electronic energy distribution on the time scale of electron-electron scattering and will especially vary if nonlinear interactions contribute to light absorption. The nonlinearity can usually be measured by varying the irradiance over only a small range around some value $E_0$ with associated temperature $T_{e,0}$, and thus the temperature-dependent and photon energy-dependent $p$ can usually take well-defined values. For the case of luminescent intraband transitions and taking into account only emission at photon energies $h\nu \gg k_B T_e$, one finds $p(h\nu) \approx h\nu/ak_B T_{e,0}$.[30]

Figure 2 more closely analyses the irradiation-dependent anti-Stokes luminescence spectra and associated nonlinearities for both the single nanoparticle site A as well as the particle cluster site C. While the total emission of each site strongly increases as a function of the irradiation, its spectral shape seems to be unaffected on first inspection. Nonetheless, from spectral measurements over limited irradiance ranges in both the low and high irradiance regime, the power-law exponent $p$ can be determined with sufficiently small error and high spectral resolution as shown in Figure 2b. In the regime of high irradiation, $p(h\nu)$ clearly displays the linear relationship with the emitted photon energy in accordance with the intraband emission model. As has been demonstrated experimentally for randomly nanostructured silver and gold surfaces,[30] the exponent depends on the approximate temperature as $p \propto T_{e,0}^{-1}$, and thus the slope of $p(h\nu)$ should increase at weaker irradiation. This is clearly not the case for the nanoparticles studied here: at low irradiance, the data instead display a much *weaker* slope that is suggestive of a constant $p = 2$. Generally, we find luminescence with similar spectral and power-law properties in both single, isolated particles as well as in particle aggregates, both for particles imaged as-prepared on bare glass surfaces as well as for particles on glass/ITO substrates before and after SEM imaging. For c.w. excitation close to the interband transition below 500 nm, contrasting claims exist on whether single nanoparticles show detectable luminescence or if the presence of a plasmonic gap mode is a necessary precondition.[12,32,33] Such gap modes can be provided both by proximate particles as well as by conductive substrates, where an image charge plasmon mode arises.[12] For the case of pulsed infrared excitation studied here, we do not find such gap modes to be a necessary precondition for luminescence. Fig 1b instead shows that luminescence efficiency varies significantly from particle to particle, with some particles emitting strongly and others showing no discernible emission. This variation indicates that either minor differences in shape can shift the longitudinal plasmon resonance sufficiently to reduce the absorption cross section or that defects and impurities play a role in the momentum selection rules governing the intraband transitions[16]. Although single particles do show luminescence, we found it impossible to study them for an extended period of time at high irradiation conditions without visible degradation of the luminescence spectrum, preventing the collection of more detailed data of the power-law behavior. On the other hand, even small particle clusters consisting of only two metal nanoparticles consistently feature strongly enhanced luminescence as well as an increased tolerance against photomodification. Pending a more extensive statistical analysis, we tentatively attribute this effect to the funneling of excitation energy from absorbing nanoparticles to emitter sites, thereby increasing the effective absorption cross-section and thus allowing luminescence to be studied at lower irradiation conditions. At the same time, we expect heat dissipation to be more effective in closely packed aggregates because of inter-particle thermal conduction, thereby reducing the risk of photothermal destruction of the emitting particle.

Figure 3 shows a series of four emission spectra in both the Stokes and anti-Stokes range for an as-prepared small cluster of metal nanoparticles on top of a bare glass surface. Again, while the spectra have similar shapes when plotted on the logarithmic scale, a range of information can be gathered from the irradiation and frequency dependence of the power-law exponent $p$. Sets of seven individual luminescence spectra were collected over one of four narrow irradiation ranges, covering levels from below 1 kW/cm² to above 10 kW/cm². Upward and downward sweeps in irradiance were considered to minimize the systematic effect of photomodification on the emission. With this approach, four sets of power-law exponents $p(h\nu)$ could be calculated for the individual irradiance ranges. In order to exclude any contribution of the substrate to the power-law exponent in the Stokes regime, spectra for the lowest irradiation range of 0.4-0.9 kW/cm² were acquired under conditions of wide-field illumination and corrected against background luminescence from the glass substrate (see the Supporting Information for details). Close to the excitation laser line between 1.3 eV and 2.0 eV, as well as under conditions of irradiation below 5 kW/cm², power-law exponents prevail which are independent of photon energy, with $p = 1$ for the Stokes and $p = 2$ for the anti-Stokes spectral range.

Such a photon-energy independent integer exponent is reminiscent of previous reports on the luminescence of gold particles and rough gold surfaces excited by c.w. or ultrafast pulsed infrared lasers,[7] where the linear luminescence response in the Stokes regime was interpreted in terms of intraband transitions and the quadratic response in the anti-Stokes regime was assigned to two-photon excited interband transitions. For irradiation levels above 5 kW/cm², the power-law exponents for luminescence below 1.3 eV and above 2.0 eV roughly follow a straight sloped line. This functionality suggests that for these irradiation levels, a unified physical mechanism could be invoked in explaining both the Stokes and anti-Stokes luminescence, with the dominant effect being a substantial irradiance-dependent blue-shift of the emission spectrum as evidenced by a linear $p(h\nu)$ functionality.

In principle, any luminescence model that involves conduction-band electronic states can be modified to include the effect of heating of the electronic energy distribution. Figure 4 sketches three popular models that have been advanced in the context of light emission from metal nanoparticles: (i) luminescent interband transitions between the *sp* conduction band and empty states in the lower-lying *d*-band;[27,28,34,35] (ii) inelastic light scattering involving transitions within the *sp* conduction band;[12,13,17] and (iii) luminescent *sp* intraband transitions.[7,30,36] The general trend in all of these models is an increased availability of high-energy transitions and a concomitant spectral blue-shift upon heating of the *sp*-band electrons. For the case of interband luminescence (i), a prior creation of holes in the *d*-band is understood to be the result of, e.g., transitions induced by two-photon absorption.[28] Luminescent transitions occur between Fermi-Dirac distributed electrons in the *sp*-band and holes near the top of the *d*-band.[37] The luminescence emitted at energy $h\nu$ under incident irradiance $E$ is then given by $\phi \propto \bar{n}_\text{F}(h\nu - \epsilon_d, T_e)$, where $\bar{n}_\text{F}(h\nu, T) = [\exp(h\nu/k_\text{B}T) + 1]^{-1}$ is the Fermi-Dirac occupation number and $\epsilon_d \approx 2.2$ eV is the energy gap[37] between $sp$ and $d$ bands at the Fermi level. The effective nonlinearity due to irradiance-induced electron heating in this approximation is $p(h\nu) = \frac{h\nu - \epsilon_d}{a\, k_\text{B}\, T_{e,0}}[1 - \bar{n}_\text{F}(h\nu - \epsilon_d, T_{e,0})]$ if $E$ is varied in a small range around $E_0$, with $T_{e,0}^a \propto E_0$.

For the case of inelastic light scattering (ii), the temperature dependence of the scattered light $\phi$ at energy $h\nu$ and incident irradiance $E$ at photon energy $h\nu_\text{inc}$ is given by $\phi \propto E|1 + \bar{n}_\text{B}(h\nu_\text{inc} - h\nu, T_e)|$, where $\bar{n}_\text{B}(h\nu, T) = [\exp(h\nu/k_\text{B}T) - 1]^{-1}$ is the Bose-Einstein occupation number.[16,17] From this relation, an effective power-law exponent $p(h\nu) = 1 + \frac{h\nu - h\nu_\text{inc}}{a\, k_\text{B}\, T_{e,0}}[1 + \bar{n}_\text{B}(h\nu - h\nu_\text{inc}, T_{e,0})]$ can be derived.

Lastly, for the case of direct intraband luminescence (iii), the temperature dependence of the luminescence emitted at energy $h\nu$ is instead described by $\phi \propto \bar{n}_\text{B}(h\nu, T_e)$ and $p(h\nu) = \frac{h\nu}{ak_\text{B}T_{e,0}}[1 + \bar{n}_\text{B}(h\nu, T_{e,0})]$.[30] Details of these calculations are given in the Supporting Information.

Figure 4 (b) shows emission spectra for an as-prepared nanoparticle cluster on glass at four irradiation levels between 5.0 kW/cm² and 11.9 kW/cm², while panel (c) shows such spectra for a similar cluster on glass/ITO. The two sites differ by an order of magnitude in brightness, but otherwise feature similar luminescence spectra. The corresponding spectrally resolved power-law exponents (panels d, e) were derived by fitting a total of seven spectra including both upward and downward sweeps in irradiance for each emission site, and are shown together with the fit error. The solid gray lines are $p(h\nu)$ model curves for the intraband luminescence model (iii) at an electronic temperature of $T_{e,0} = 4000$ K (panel c) and 3700 K (panel d). For photon energies above 1 eV the models are only sensitive to the product $a \cdot T_{e,0}$. Since the data in the spectral region below 1 eV is insufficient, a reliable estimate of $a$ is impossible so that the free-electron value $a = 2$ was tentatively used for the present analysis. Except for the spectral region close to the excitation laser and luminescence intensities near the detection threshold, the model reproduces the experimental data rather well. For high photon energies, $p$ converges towards a linear relationship $p(h\nu) = \frac{h\nu}{a\, k_B\, T_{e,0}}$ with slope 1.45 eV$^{-1}$ (panel c) and 1.57 eV$^{-1}$ (panel d). Both inelastic light scattering (ii) as well as interband luminescence (i) allow for the same linear functionality in the regime of high photon energies, although these are shifted either by the energy of the incident photons as $p(h\nu) = \frac{h\nu - h\nu_{\text{inc}}}{a\, k_B\, T_{e,0}}$ in the case of inelastic scattering, or by the interband energy gap $\epsilon_d$ as $p(h\nu) = \frac{h\nu - \epsilon_d}{a\, k_B\, T_{e,0}}$ in the case of interband luminescence. We plot both model curves at the same temperature as the intraband luminescence model in order to match the high-energy slope (dashed and dot-dashed lines). Neither of these model curves are able to correctly follow the data points. In order to reach power-law exponents similar to the experimental ones in the high-energy regime, lower electronic temperatures are needed, resulting in a much stronger slope of $p(h\nu)$ which is not supported by the data. An example of a fit to the inelastic scattering model at lower $T_{e,0}$ is shown as dotted lines (ii').

As experimentally demonstrated earlier for randomly structured silver surfaces,[30] the slope of $p(h\nu)$ varies inversely with the effective electronic temperature, in apparent contradiction to the experimental observation of an increased slope at higher irradiance. Indeed, we found it impossible to study this characteristic of the thermal emission component on its own, given that a mixture of thermal and non-thermal emission is observed over most of the irradiation range. At even higher irradiation one would expect the thermal emission component to dominate, but any study of this regime is hindered by the onset of strong photothermal degradation.

We have shown that ultrafast pulsed infrared excitation of gold nanoparticles induces spectrally broad-band luminescence that covers the visible to near-infrared spectrum. It occurs in both single nanoparticles as well as small clusters of such particles, on both conductive as well as non-conductive surfaces and is thus not related to the presence of a plasmonic gap mode but rather to the particles themselves. Although care must be taken to correct for background emission from the substrate, both the spectral power law analysis as well as the low-irradiance wide-field measurements clearly show the emission to be distinct from any glass-related bulk effects that occur due to through-substrate excitation and observation. The same results are obtained for nanoparticles deposited directly on glass as well as on top of an ITO layer. The observed emission therefore cannot be attributed to the substrate volume within the plasmonic near field, directly underneath the particle. While for single nanoparticles light emission is not detectable until just beneath the photomodification threshold, it is more pronounced and stable in small particle clusters. This is likely due to an increased absorption cross-section and funneling of excitation energy through plasmonic coupling to proximate particles,[38] which in turn promotes electronic heating at lower irradiance and thus lower associated heating of the lattice. Under conditions of low irradiance, the luminescence depends linearly on excitation power in the Stokes region, and nonlinearly in the anti-Stokes region. The linear power dependence is suggestive of the reported signatures of electronic Raman scattering under c.w. laser irradiation.[13,14,17] On the other hand, the square power-law in the anti-Stokes region rather suggests two-photon absorption-related interband transitions occurring without significant electronic heating.[26,27] It is not clear though, why such two-photon luminescence would only impact the anti-Stokes side of the spectrum, while a linear irradiance dependence dominates on the Stokes side. Upon increasing the irradiance to levels above several kW/cm², the effective luminescence nonlinearity is altered in a characteristic way that is directly identified in a spectrally-resolved power-law analysis. The linear dependence of the effective nonlinearity on photon energy $p \propto h\nu$ suggests the emergence of an irradiation-induced heating mechanism, and can be directly explained by the increased availability of high-energy electronic states within the conduction-band electron gas. Effective electronic temperatures reach values of 4000 K, in agreement with recent theoretical predictions for metallic spheres of 10 nm diameter upon 10 fs pulsed-laser excitation under similar fluence conditions.[39]

The model used here to understand such heating effects is obviously simplified in that it does not take into account the real band structure nor the time evolution of occupied electronic states. Nonetheless, it serves to illustrate the effect of introducing heating-related effective nonlinearity to the particle's luminescence response. When combined with a simplified description of three of the most common models of nanoparticle luminescence, we find that only intraband luminescence is fully compatible with the observed combination of the absolute values of the exponent of nonlinearity $p$ as well as the slope of its spectral dependence. It appears feasible to disentangle possible contributions to emission in the low-irradiance regime from electronic Raman scattering and intraband luminescence in, e.g., excitation wavelength-dependent studies. It is also conceivable that heating-induced changes to the effective dielectric response alter the particle's plasmon resonance in a way to contribute to distortions of the observed nonlinearities. The effect would be different for particles having their resonance below or above the laser wavelength. One would need excitation wavelength-dependent studies of the power-law spectra to further clarify these points, but such measurements are hampered by the strong variation of the particle's absorption cross section when tuning with the laser wavelength. A viable approach to the problem has recently been shown with concurrent measurements of the luminescence and the elastic scattering spectrum, which allows for an intrinsic correction.[36] We consider the contribution of heating induced shifts of the plasmon resonance likely to be small, though, given that spectral power-law behavior very similar to the high-irradiance regime presented here has been found for randomly structured gold surfaces with no defined plasmon-resonance spectral position.[30]

The spectrally-resolved nonlinearity analysis of metal luminescence discussed here provides a useful tool to help disentangle the different processes contributing to nanoparticle luminescence phenomena. The illustration of a power-dependent transition from $p = $ const. to $p \propto h\nu$ on one and the same particle helps explain the surprising array of mutually seemingly contradicting observations concerning luminescence nonlinearity documented in past publications. Even single particles apparently can give rise to luminescence that can be interpreted equally in terms of linear, two-photon or non-integer nonlinear interactions, depending on the spectral range and irradiance level studied.

**Acknowledgements**

We are indebted to Kaiqiang Lin (Xiamen University) for helpful discussions.

**Funding Information**

The authors are indebted to the DFG for funding through GRK 1570 and to the European Research Council (ERC) for funding through the Starting Grant MolMesON (305020).

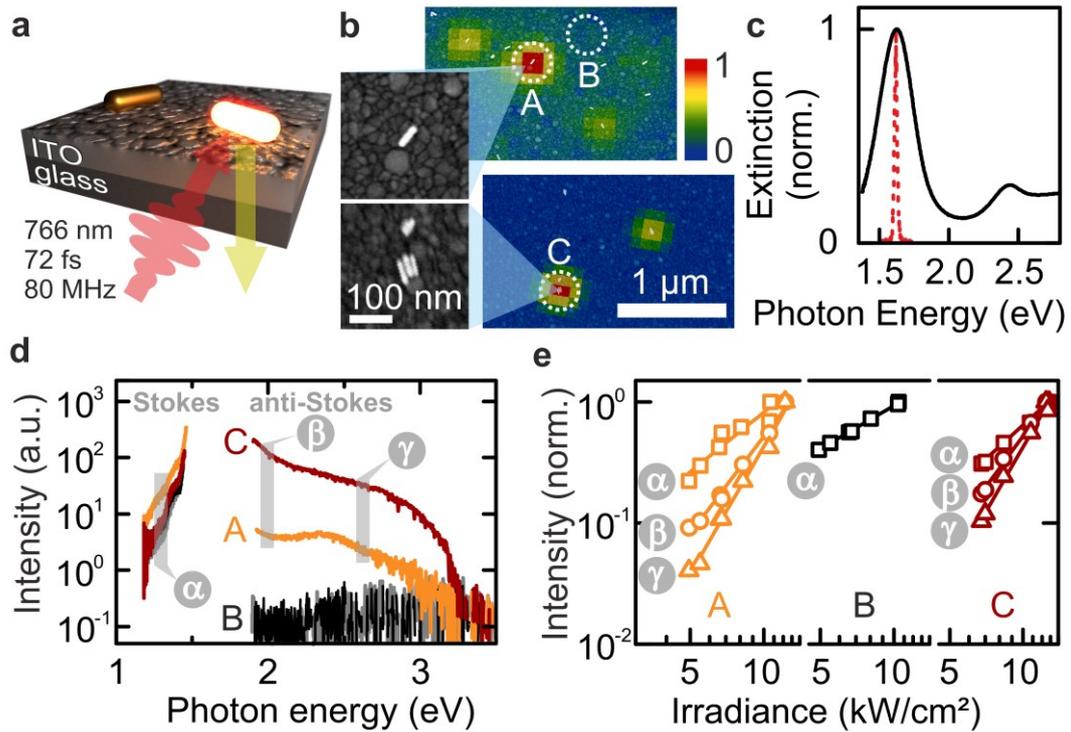

**Figure 1.** Photoluminescence of metal nanoparticles under focused femtosecond infrared laser excitation. (a) Schematic of gold nanoparticles on the surface of a glass substrate covered by a layer of conductive indium-tin oxide (ITO), with through-glass collimated or focused laser excitation (red) and PL emission (orange). (b) False-color normalized anti-Stokes PL maps under circularly polarized wide-field excitation, superimposed on a SEM topography image of the sample. Individual sites are marked A (single particle), B (no particle), and C (small particle cluster), with SEM zooms to the left showing details for A and C. (c) Extinction of gold nanoparticles dispersed in water (black) and excitation laser spectrum (red). (d) Emission spectra measured at 10.5 kW/cm² focused excitation of site A (orange), site B (black), site C (crimson) and from a blank glass/ITO substrate without any nanoparticles (grey). (e) Irradiance dependence of emission intensity from sites A, B, and C integrated over narrow spectral regions marked $\alpha$, $\beta$ and $\gamma$ in panel (d). Data points include downward and upward sweeps of irradiance to minimize the effect of photodegradation, and the lines show linear least-squares fits.

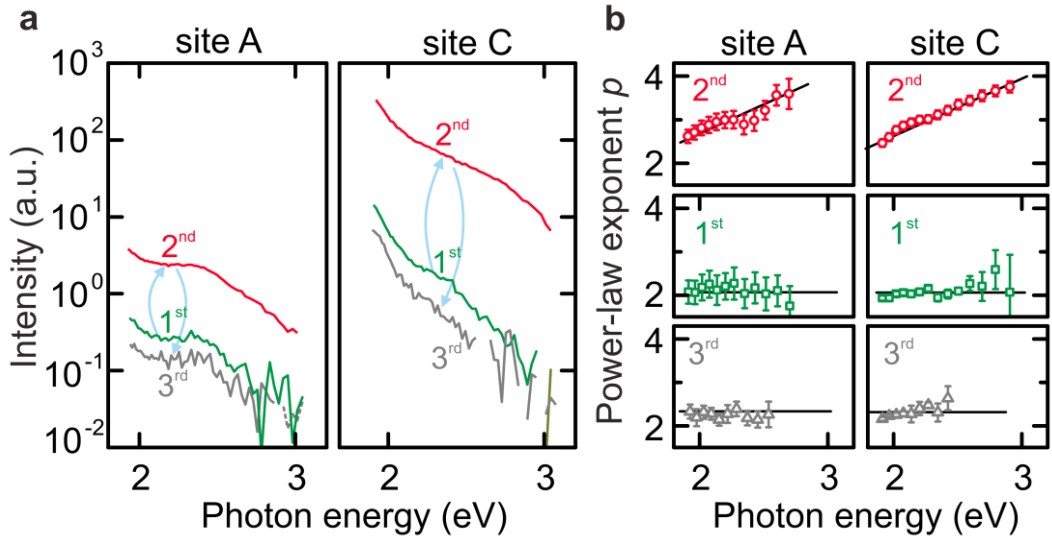

**Figure 2.** Irradiance-dependent anti-Stokes emission spectra from emissive sites A (single particle) and C (particle aggregate), and analysis of their power-law behaviour. (a) Emission spectra for sites A (left) and C (right) at high (10.8 kW/cm², red line) and low irradiance (3.5 kW/cm² for site A, 1.4 kW/cm² for site C). The low irradiance spectra were recorded before (green) and after (grey) obtaining the high irradiance spectrum. The actual measurement order 1st-3rd of the spectra is indicated by arrows. (b) Spectrally resolved power-law exponents $p$ corresponding to panel (a), each value being calculated for a narrow spectral region and from seven individual emission intensities measured over a limited range of irradiance as illustrated in Fig 1e. The irradiance ranges covered by the sweeps for site A were 3.5-5.4 kW/cm² (1st and 3rd data sets) and 5.0-10.6 kW/cm² (2nd data set). For site C, the irradiance ranges were 0.87-1.43 kW/cm² (1st and 3rd data sets) and 6.8-12.7 kW/cm² (2nd data set). The sweeps included a down- and upward sweep in irradiance and the error bars relate to the standard error of a linear fit of Fig. 1e.

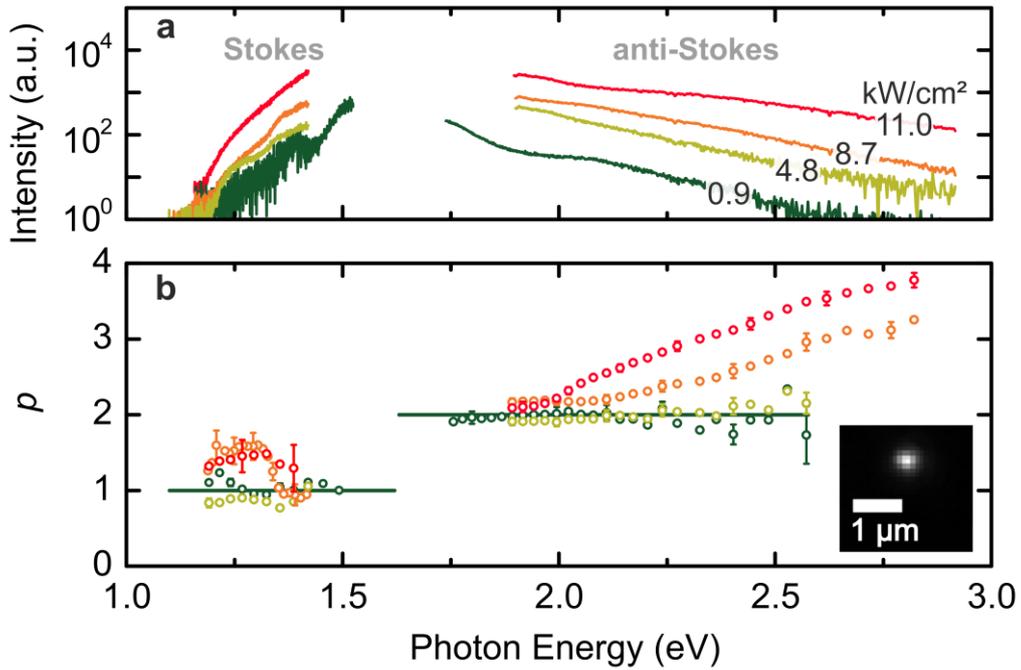

**Figure 3.** Gradual crossover of spectrally resolved power-law functionalities. (a) Emission spectra of a small particle aggregate on a glass surface for increasing irradiation (green: 0.9 kW/cm², light green: 4.8 kW/cm², orange: 8.7 kW/cm², red: 11.0 kW/cm²). Scattered light is cut off by appropriate filters. The spectrum for 0.9 kW/cm² irradiation is measured using wide-field illumination, all others using a focused excitation beam. (b) Power-law exponents $p$, each derived from a series of seven measurements in up- and downward sweeps of laser power over a limited range of irradiation (green: 0.4-0.9 kW/cm², light green: 1.8-4.9 kW/cm², orange: 2.6-8.7 kW/cm², red: 4.5-11.0 kW/cm²). For clarity, error bars derived from the fit are included only for a subset of the data points. Inset: image of the diffraction-limited emission spot in the anti-Stokes spectral region.

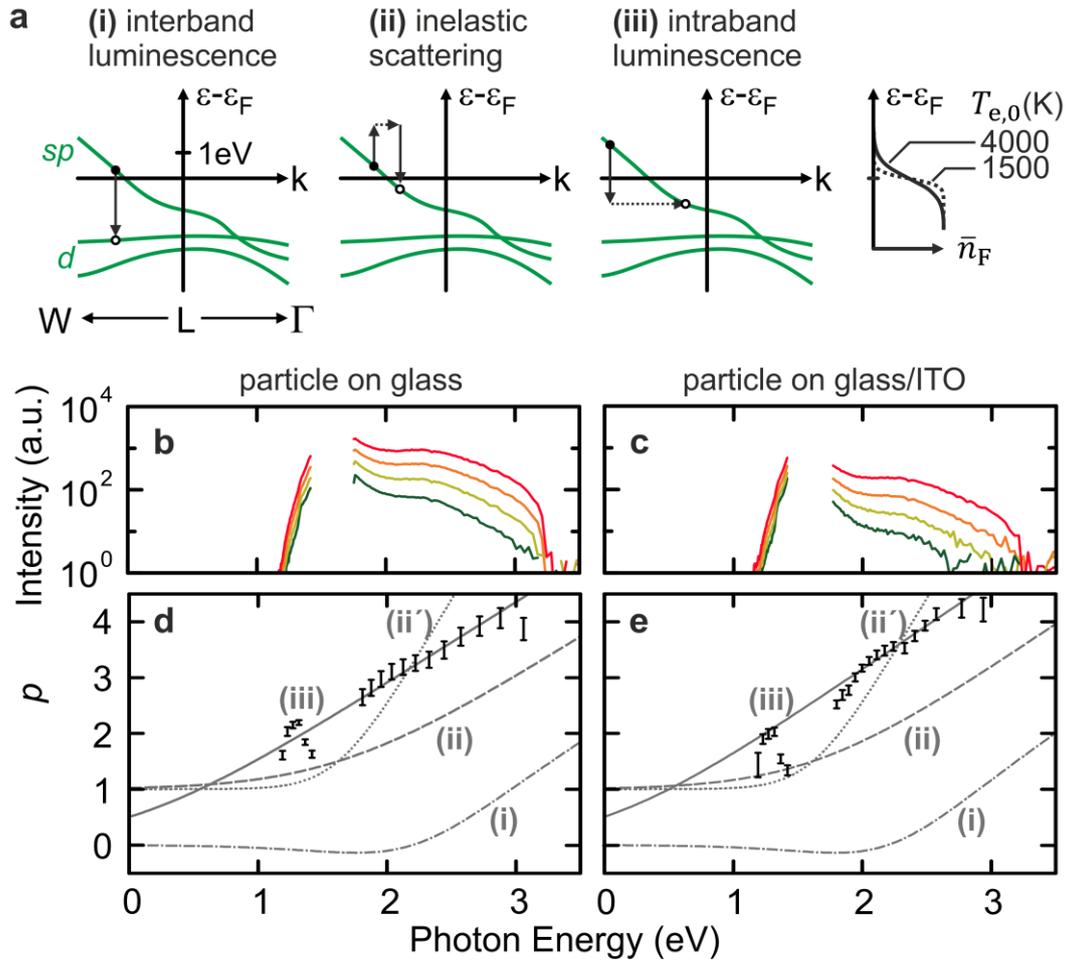

**Figure 4.** Comparison of spectrally resolved power-law data to model functionalities for two individual particle aggregate sites on different substrates. (a) Schematic band structure near the L symmetry point, with three types of luminescence processes indicated. The Fermi-Dirac distribution for the heated sp-band electron gas is shown on the right. (b-c) Spectra recorded for each single-particle aggregate site at an irradiation of 5.0 kW/cm² (green), 6.9 kW/cm² (light green), 8.9 kW/cm² (orange), 11.9 kW/cm² (red). (d-e) Corresponding spectrally resolved power-law exponents $p$ derived from the measured spectra including fit errors (black). Model curves calculated from the interband luminescence model (curve i, dot-dashed) at $T_{e,0} = 4000$ K (panel c) and $T_{e,0} = 3700$ K (panel d) as well as for for inelastic light scattering (curve ii, dashed) and intraband luminescence (curve iii, solid) at the same electronic temperatures. For comparison, model curves for inelastic light scattering at a temperature of $T_{e,0} = 1500$ K are included as well (curve ii', dotted). For all curves, the thermal exponent $a = 2$ was chosen.